\documentclass{article}
\usepackage{spconf,amsmath,graphicx}
\usepackage{color}
\usepackage{booktabs}
\usepackage{multirow}
\usepackage{soul}
\usepackage{amsfonts,amssymb}
\usepackage[labelformat=simple]{subcaption}
\usepackage{hyperref}

\title{Transformer-based End-to-End Speech Recognition with Local Dense Synthesizer Attention}

\name{Menglong Xu, Shengqiang Li, Xiao-Lei Zhang
}

\address{CIAIC, School of Marine Science and Technology, Northwestern Polytechnical University, China}

\email{\{mlxu, shengqiangli\}@mail.nwpu.edu.cn, xiaolei.zhang@nwpu.edu.cn}

\begin{document}

\maketitle

\begin{abstract}
  Recently, several studies reported that dot-product self-attention (SA) may not be indispensable to the state-of-the-art Transformer models.
  Motivated by the fact that dense synthesizer attention (DSA), which dispenses with dot products and pairwise interactions, achieved competitive results in many language processing tasks, in this paper, we first propose a DSA-based speech recognition, as an alternative to SA. To reduce the computational complexity and improve the performance, we further propose local DSA (LDSA) to restrict the attention scope of DSA to a local range around the current central frame for speech recognition. Finally, we combine LDSA with SA to extract the local and global information simultaneously. Experimental results on the Ai-shell1 Mandarin speech recognition corpus show that the proposed LDSA-Transformer achieves a character error rate (CER) of 6.49\%, which is slightly better than that of the SA-Transformer. Meanwhile, the LDSA-Transformer requires less computation than the SA-Transformer. The proposed combination method not only achieves a CER of 6.18\%, which significantly outperforms the SA-Transformer, but also has roughly the same number of parameters and computational complexity as the latter.
  The implementation of the multi-head LDSA is available at https://github.com/mlxu995/multihead-LDSA
\end{abstract}


\section{Introduction}\label{sec:introduction}

  \begin{figure*}[t]
    \begin{subfigure}{.36\linewidth}
      \centering
      \includegraphics[width=3.6cm]{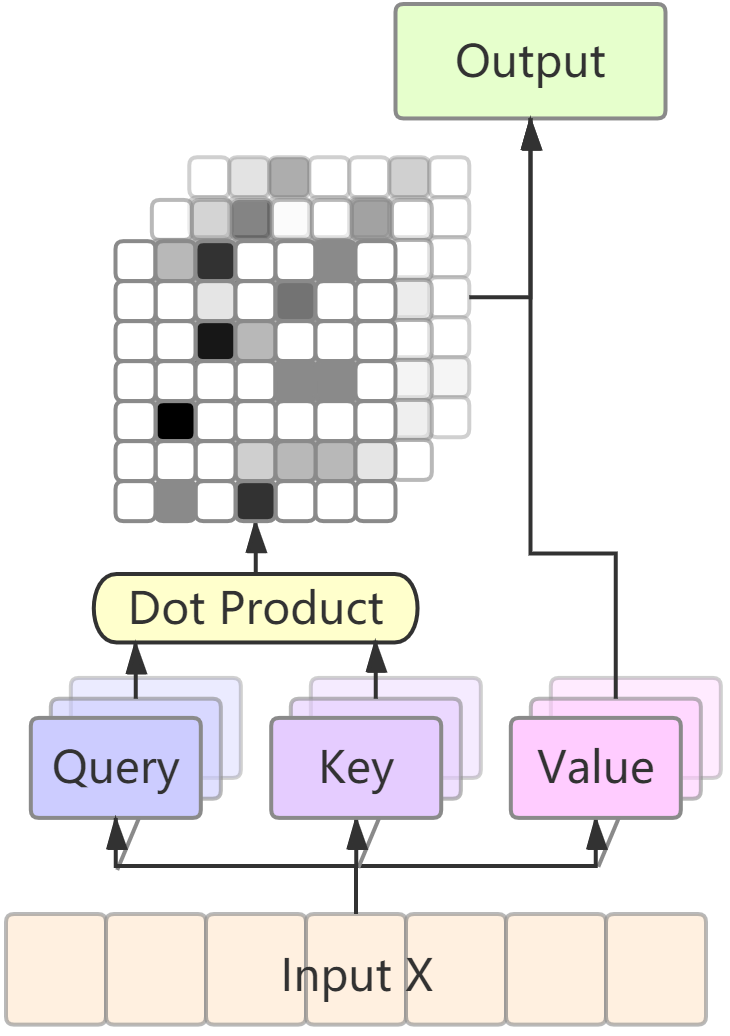}
      \caption{Self-attention (with 3 heads)}
      \label{fig/attention_a}
    \end{subfigure}\hspace{-20mm}
    \hfill
    \begin{subfigure}{.23\linewidth}
      \centering
      \includegraphics[width=3.6cm]{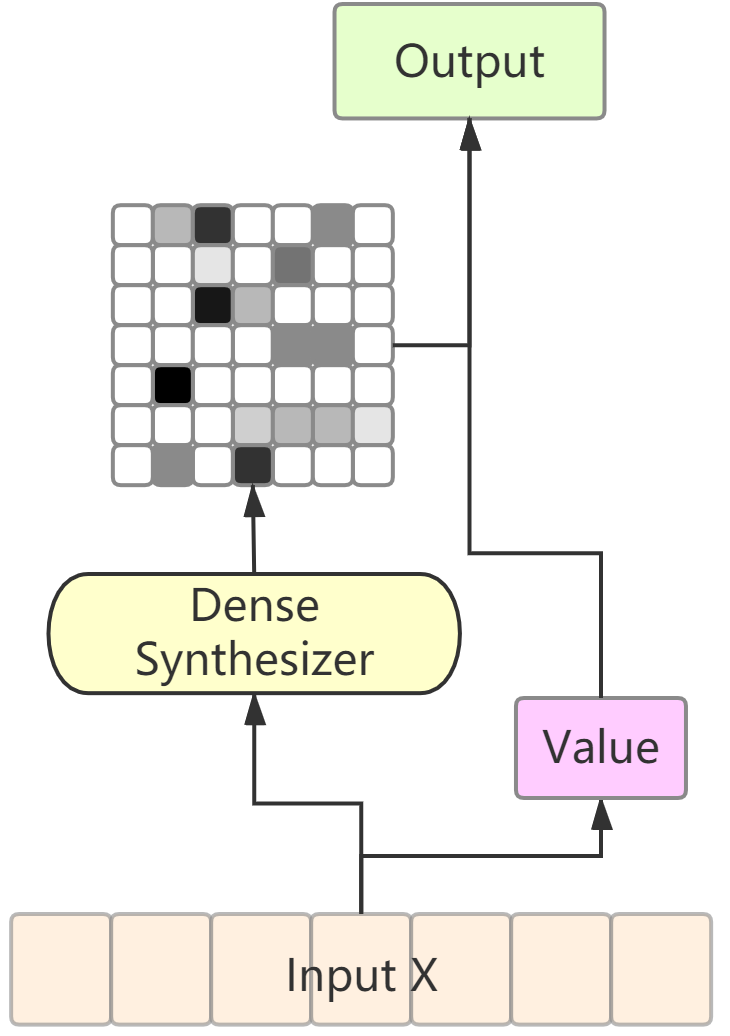}
      \caption{Dense synthesizer attention}
      \label{fig/attention_b}
    \end{subfigure}\hspace{-20mm}
    \hfill
    \begin{subfigure}{.36\linewidth}
      \centering
      \includegraphics[width=3.6cm]{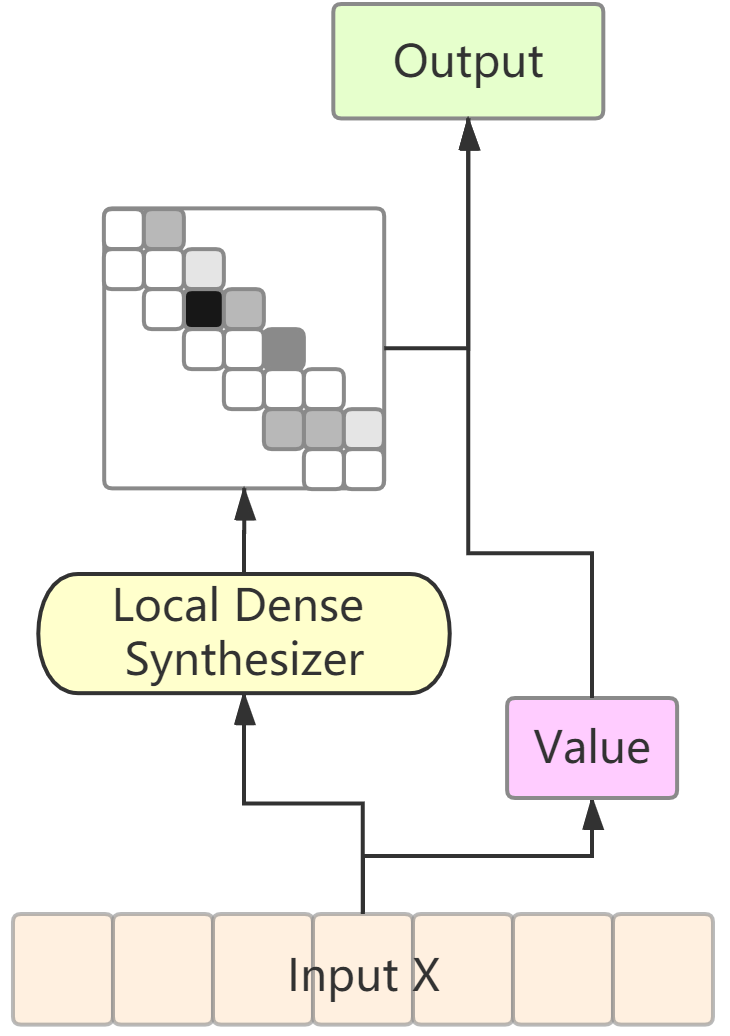}
      \caption{Local dense synthesizer attention}
      \label{fig/attention_c}
    \end{subfigure}
    \caption{Architecture of different attention mechanisms.}
    \label{fig:attentions}
  \end{figure*}

  In recent years, end-to-end (E2E) automatic speech recognition (ASR) \cite{amodei2016deep,chan2016listen,battenberg2017exploring,
  watanabe2017hybrid,dong2018transformer,zhou2018syllable} has been widely studied in the ASR community due to its simplified model structure as well as its simple training and inference pipelines.
  Among various E2E models, Transformer-based ASR \cite{karita2019comparative,yeh2019transformer,moritz2020streaming,wang2020transformer,zhang2020transformer} has received more and more attention for its high accuracy and efficient training procedure.
  The core component of the state-of-the-art Transformer-based models is a so-called self-attention mechanism \cite{vaswani2017attention}, which uses dot products to calculate attention weights.
  Although the content-based dot-product self-attention is good at capturing global interactions, it makes the computational complexity of the self-attention (SA) layer be quadratic with respect to the length of the input feature.

  Therefore, there is a need to reduce the complexity of the SA layer. Fortunately, several recent studies in natural language processing simplified the expensive dot-product self-attention \cite{chiu2018monotonic,wu2019lite,wu2018pay, raganato2020fixed,tay2020synthesizer}. Specifically,
  In \cite{wu2018pay}, SA was replaced with a so-called dynamic convolution. It uses an additional linear layer to predict normalized convolution weights dynamically at each convolution step.
  %\textcolor[rgb]{0.00,0.50,1.00}{In \cite{chiu2018monotonic}, Monotonic Chunkwise Attention was proposed to make the decoding process linear-time.}
  In \cite{raganato2020fixed}, Raganato \textit{et al.} replaced all but one attention heads with simple fixed (non-learnable) attention patterns in Transformer encoders.
  In \cite{tay2020synthesizer}, Tay \textit{et al.} proposed dense synthesizer attention (DSA), which uses two feed-forward layers to predict the attention weights. Compared to SA, DSA completely dispenses with dot products and explicit pairwise interactions. It achieves competitive results with SA across a number of language processing tasks.

  However, it is not easy to replace SA by DSA in ASR.
  First, the length of the attention weights predicted by DSA is fixed. If we apply DSA directly to ASR, then the spectrogram of each utterance has to be padded to the length of the longest utterance of the training corpus, which unnecessarily consumes quite long time and large storage space.
  Moreover, the length of the feature in an ASR task is much longer than that in a language model.
  Predicting attention weights directly for such a long spectrogram results in a significant increase of errors.
  In addition, like SA, DSA still does not have the ability to extract fine-grained local feature patterns.

  In this paper, we propose local dense synthesizer attention (LDSA) to address the aforementioned three problems simultaneously.
  In LDSA, the current frame is restricted to interacting with its finite neighbouring frames only. Therefore, the length of the attention weights predicted by LDSA is no longer the length of the longest utterance. It is a fixed length controlled by a tunable \textit{context width}. LDSA not only reduces the storage and computational complexity but also significantly improves the performance.

  To evaluate the effectiveness of the LDSA-Transformer, we implemented the DSA-Transformer, LDSA-Transformer, and the combination of the LDSA and SA for ASR, where we denote the combined model as \textit{hybrid-attention} (HA) Transformer. Experimental results on the Ai-shell1 Mandarin dataset show that
  the LDSA-Transformer achieves slightly better performance with less computation than the SA-Transformer.
  In addition, HA-Transformer achieves a relative character error rate (CER) reduction of 6.8\% over the SA-Transformer with roughly the same number of parameters and computation as the latter.

  The most related work of LDSA is \cite{fujita2020attention}, in which Fujita \textit{et al.} applied dynamic convolution \cite{wu2018pay} to E2E ASR.
  However, the method \cite{fujita2020attention} is fully convolution-based. It does not adopt the SA structure.
  On the contrary, our model adopts the SA structure instead of the convolution structure.
  In addition, we combine the proposed LDSA with SA by replacing the convolution module in the convolution-augmented Transformer with LDSA, so as to further model the local and global dependencies of an audio sequence simultaneously.

  %The remainder of the paper is organized as follows. Section \ref{}  introduces the proposed XXX. Section \ref{} presents the experimental setup and results. Section \ref{} concludes the paper.

%\section{Relative work}

\section{algorithm description}\label{sec:algorithm}

  In this section, we first briefly introduce the classic dot-product self-attention and its variant---DSA, and then elaborate the proposed LDSA.
  \subsection{Dot-product self-attention}
   The SA in transformer usually has multiple attention heads. As illustrated in Fig. \ref{fig/attention_a}, suppose the multi-head SA has $h$ heads. It calculates the scaled dot-product attention $h$ times and then concatenates their outputs. A linear projection layer is built upon the scaled dot-product attention, which produces the final output from the concatenated outputs.
   Let $\mathbf{X}\in \mathbb{R}^{T\times d}$ be an input sequence, where $T$ is the length of the sequence and $d$ is the hidden size of the SA layer. Each scaled dot-product attention head is formulated as:
    \vspace{-0.15cm}
    \begin{equation}
    \text{Attention}(\mathbf{Q}_i,\mathbf{K}_i,\mathbf{V}_i)=
    \text{Softmax}\left(
    \frac{\mathbf{Q}_i \mathbf{K}_i^{\mathrm{T}}}
    {\sqrt{d_{\mathrm{k}}}}\right) \mathbf{V}_i \\
    \end{equation}
    \vspace{-0.15cm}
    with
    \vspace{-0.15cm}
    \begin{equation}
    \mathbf{Q}_i=\mathbf{X W}^{\mathrm{Q}_i},\
    \mathbf{K}_i=\mathbf{X W}^{\mathrm{K}_i},\
    \mathbf{V}_i=\mathbf{X W}^{\mathrm{V}_i}
    \end{equation}
    where $\mathbf{W}^{\mathrm{Q}_i},\mathbf{W}^{\mathrm{K}_i},
    \mathbf{W}^{\mathrm{V}_i}
    \in\mathbb{R}^{d\times d_\mathrm{k}}$ denote learnable
    projection parameter matrices for the $i$-th head, $d_\mathrm{k}=d/h$ is the dimension of the feature vector for each head.
    The multi-head SA is formulated as:
    \begin{equation}
    \begin{split}
    \text{MultiHead}(\mathbf{Q},\mathbf{K},\mathbf{V})=
    \text{Concat}\left(
    \mathbf{U}_1,\cdots,\mathbf{U}_h
    \right)
    \mathbf{W}^\mathrm{O}\\
    \end{split}
    \end{equation}
    \vspace{-0.15cm}
    where
    \vspace{-0.15cm}
    \begin{equation}
    \begin{split}
    \ \mathbf{U}_i=
    \text{Attention}\left(
    \mathbf{X}\mathbf{W}^{\mathrm{Q}_{i}},
    \mathbf{X}\mathbf{W}^{\mathrm{K}_{i}},
    \mathbf{X}\mathbf{W}^{\mathrm{V_{i}}}\right) \\
    \end{split}
    \end{equation}
    and $\mathbf{W}^\mathrm{O}\in\mathbb{R}^{d\times d}$ is the weight matrix of the linear projection layer.

  \subsection{Dense synthesizer attention}
     As illustrated in Fig. \ref{fig/attention_b}, the main difference between DSA and SA is the calculation method of the attention weights. Dense synthesizer attention removes the notion of query-key-values in the SA module and directly synthesizes the attention weights.
     In practice, DSA adopts two feed-forward layers with ReLU activation to predict the attention weights,
     which is formulated as:
     \begin{equation}\label{equ:B}
     \mathbf{B}=\text{Softmax}(
     \sigma_\mathrm{R}(\mathbf{X}\mathbf{W}_1)
     \mathbf{W}_2)
     \end{equation}
     where $\sigma_\mathrm{R}$ is the ReLU activation function, and $\mathbf{W}_1\in\mathbb{R}^{d\times d}$ and $\mathbf{W}_2\in\mathbb{R}^{d\times T}$ are learnable weights.
     The output of DSA is calculated by:
     \begin{equation}
     \text{DSA}(\mathbf{X})=\mathbf{B}
     (\mathbf{X}\mathbf{W}_3)\mathbf{W}^\mathrm{O}
     \end{equation}
      with $\mathbf{W}_3\in\mathbb{R}^{d\times d}$.

  \subsection{Proposed local dense synthesizer attention}
     Motivated by convolutional neural networks, we propose LDSA to address the weaknesses of DSA.
     LDSA restricts the current frame to interact with its neighbouring frames only.
     As illustrated in Fig. \ref{fig/attention_c}, it defines a hyper-parameter $c$, termed as \textit{context width}, to control the length of the predicted attention weights, and then assign the synthesized attention weights to the current frame and its neighboring frames, where $c=3$ in Fig. \ref{fig/attention_c}.
     Attention weights for the other frames outside the context width will be set to 0.
     The calculation method of $\mathbf{B}$ in LDSA is the same as that in DSA. However, its time and storage complexities are reduced significantly, due to the fact that $\mathbf{W}_2 \in \mathbb{R}^{d\times c}$ in LDSA.
     The output of LDSA is calculated by:
     \begin{eqnarray}
      \mathbf{V}=\mathbf{X}\mathbf{W}_3\ \ \ \ \ \ \ \ \ \ \ \\
      \mathbf{Y}_t=\sum_{j=0}^{c-1}
      \mathbf{B}_{t,j}\mathbf{V}_{t+j-\lfloor\frac{c}{2}\rfloor}\\
      \text{LDSA}(\mathbf {X})=\mathbf{Y}\mathbf{W}^\mathrm{O}\ \ \
     \end{eqnarray}
     Both DSA and LDSA can be easily extended to a multi-head form in a similar way with the dot-product self-attention.
     %The efficient implementation of our multi-head LDSA can be found at https://github.com/mlxu995/multihead-LDSA.

%\section{Model implementation}
%    This section first describes the baseline model---SA-Transformer, and then presents the proposed LDSA-Transformer.

\section{Model implementation}
    This section first describes the baseline model, and then presents the proposed models.
    \subsection{Baseline model: SA-Transformer}
\begin{figure}[t]
      \centering
      \centerline{\includegraphics[width=6.5cm]{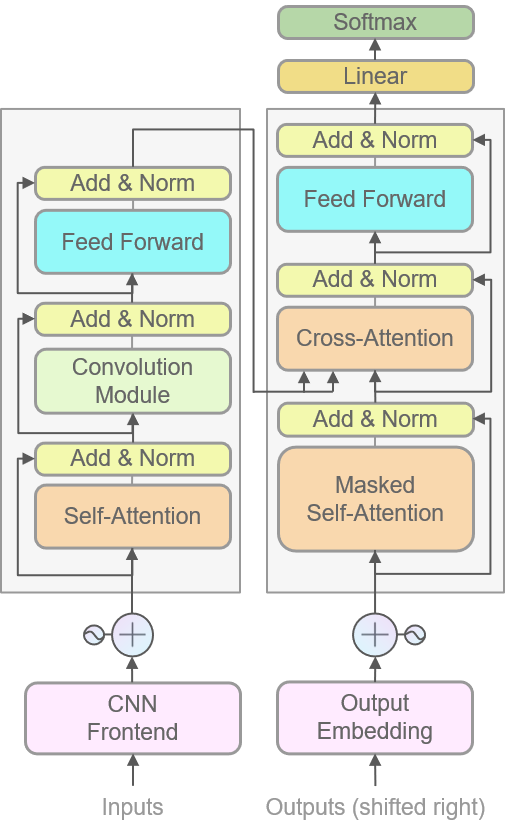}}
      %\vspace{1.5cm}
     %\caption{Transformer model architecture.}
     \caption{The model architecture of the SA-Transformer.}
     \label{fig/F2}
     \end{figure}

     The SA-Transformer is an improved Speech-transformer \cite{dong2018transformer}.
     As shown in Fig. \ref{fig/F2}, it consists of an encoder and a decoder.
     The encoder is composed of a convolution frontend and a stack of $N=12$ identical encoder sub-blocks, each of which contains a SA layer, a convolution layer\footnote{Unlike Conformer \cite{gulati2020conformer}, we only added the convolution layer without the relative positional encoding.} and a position-wise feed-forward layer.
     For the convolution frontend, we stack two $3\times3$ convolution layers with stride 2 for both time dimension and frequency dimension to conduct down-sampling on the input features.
     The decoder is composed of an embedding layer and a stack of $M=6$ identical decoder sub-blocks.
     In addition to the position-wise feed-forward layer, the decoder sub-block contains two SA layers performing multi-head attention over the embedded label sequence and the output of the encoder respectively.
     The output dimension of the SA and feed-forward layers are both 320. The number of the attention heads in each SA layer is 4.
     Note that we also add residual connection and layer normalization after each layer in the sub-blocks.

    \subsection{Proposed LDSA-Transformer}
     The LDSA-Transformer has the same decoder as the baseline model. It replaces the self-attention mechanism in the encoder of the SA-Transformer with LDSA. The number of heads of LDSA is set to 4. The other layers in the encoder of the LDSA-Transformer are the same as the baseline model. As for the DSA-Transformer, it just changes LDSA in the LDSA-transformer to DSA.

\subsection{Proposed HA-Transformer}
     The HA-Transformer is a combination of SA and the proposed LDSA. Different from the additive operation as \cite{tay2020synthesizer} did,
     we combine them in a tandem manner since that LDSA is able to extract fine-grained local patterns, which is similar to \cite{gulati2020conformer}. The difference between the HA- and SA-Transformers is that the HA-Transformer uses LDSA to replace the convolution layers in the baseline model, leaving the rest of the SA-Transformer unchanged.
     For a fair comparison, we set $c=15$ in HA-Transformer, which equals to the size of the convolution kernel in SA-Transformer.

\section{Experiments}\label{sec:experiments}
  \subsection{Experimental setup}
   We evaluated the proposed models on a publicly-available Mandarin speech corpus Aishell-1 \cite{bu2017aishell}, which contains about 170 hours of speech recorded from 340 speakers.
   We used the official partitioning of the dataset, with 150 hours for training, 20 hours for validation, and 10 hours for testing.
   For all experiments, we used 40-dimension Mel-filter bank coefficients (Fbank) features as input.
   The frame length and shift was set to 25 ms and 10 ms respectively.
   For the output, we adopted a vocabulary set of 4230 Mandarin characters and 2 non-language symbols, with the 2 symbols denoting unknown characters and the start or end of a sentence respectively.

   We used Open-Transformer\footnote{https://github.com/ZhengkunTian/OpenTransformer} to build our models.
   For the model training, we used Adam with Noam learning rate schedule (25000 warm steps) \cite{vaswani2017attention} as the optimizer. We also used SpecAugment \cite{park2019specaugment} for data augmentation.
   After 80 epochs training, the parameters of the last 10 epochs were averaged as the final model.
   During inference, we used a beam search with a width of 5 for all models.
   For the language model, we used the default setting of Open-Transformer, and integrated it into beam search by shallow fusion \cite{kannan2018analysis}. The weight of the language model was set to 0.1 for all experiments.

% \subsection{Effect of the context width of LDSA}
  \subsection{Results}
   \begin{figure}[t]
    \centering
      \centerline{\includegraphics[width=7.5cm]{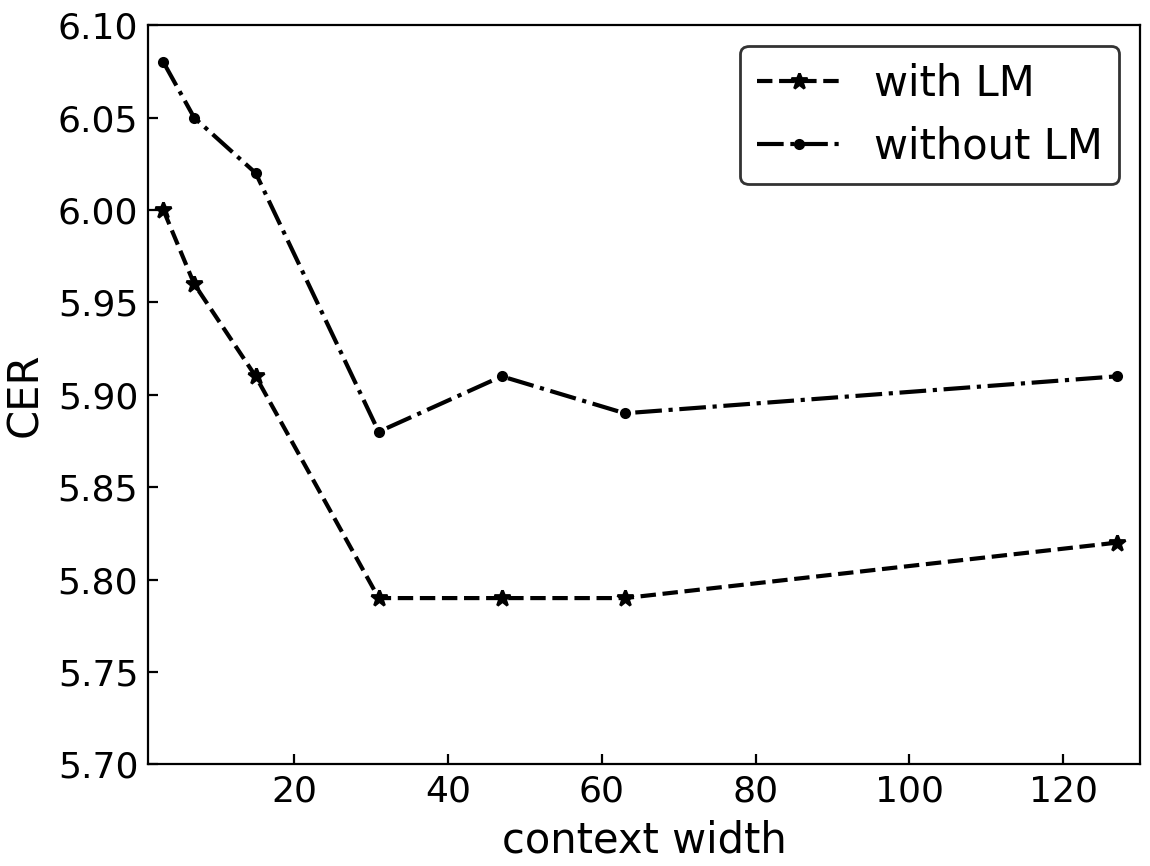}}
    % \vspace{1.5cm}
      \caption{Effect of the context width of LDSA on performance.}
    \label{fig:contex}
   \end{figure}

   We first investigated the effect of the context width $c$ of LDSA in the encoder on the development (Dev) set of Alshell-1, where we fixed the size of the convolution kernel in all experiments. Figure \ref{fig:contex} shows the CER curve of the model with respect to $c$.
   From the figure, we see that the CER first decreases, and then becomes stable with the increase of $c$.
   Based on the above finding, we set $c$ to 31 in all of the following comparisons.

% \subsection{Comparison between different attention mechanisms} \label{sec:compare_a}
   Then, we compared the attention mechanisms mentioned in Section \ref{sec:algorithm}.
   Table \ref{tab:compare_b} lists the CER and complexity of the attention mechanisms.
   From the table, we see that the LDSA-Transformer significantly outperforms the DSA-Transformer, and achieves a slightly lower CER than the SA-Transformer, which demonstrates the effectiveness of the LDSA-Transformer. We also see that the computational complexity of the LDSA scales linearly with $T$, which is lower than the SA and DSA.
   Finally, the HA-Transformer
   achieves the best performance among all comparison methods. Particularly, it achieves a relative CER reduction of 6.8\% over the SA-Transformer, which demonstrates that the LDSA performs better than the convolution operation in extracting local features.
   \begin{table}[t]
      \caption{Comparison of models with different attention mechanisms on the test set. ($T$ is the length of input feature, $c$ is the context width.)}
      \vspace{6pt}
      \label{tab:compare_b}
      \centering
      \scalebox{0.88}{
       \begin{tabular}{c c c c c c}
        \toprule
        \multirow{2}{*}{\textbf{Method}} & &
        \multirow{2}{*}{\textbf{Complexity}} & &
        \multicolumn{2}{c}{\textbf{CER}} \\
        & & & &
        \textbf{without LM} &
        \textbf{with LM} \\
        \midrule
        SA && $\mathcal{O}(T^{2})$ && 6.83 & 6.63 \\
        DSA && $\mathcal{O}(T^2)$ && 7.52 & 7.26 \\
        LDSA && $\mathcal{O}(Tc)$ && 6.65 & 6.49 \\
        HA && $\mathcal{O}(T(T+c))$ && 6.38 & 6.18 \\
        \bottomrule
       \end{tabular}
       }
   \end{table}

 % \subsection{Comparison with the representative ASR systems}
   To further investigate the effectiveness of the proposed models, we compared them with several representative ASR systems, which are the TDNN-Chain \cite{povey2016purely}, Transducer \cite{tian2019self}, and LAS \cite{shan2019component} in Table \ref{table:compare_o}.
   From the table, we find that the Transformer-based models outperform the three comparison systems \cite{povey2016purely,tian2019self,shan2019component}.
   Among the Transformer-based models, LDSA-Transformer achieves slightly better performance than the SA-Transformer. The HA-Transformer achieves a CER of 6.18\%, which is significantly better than the other models.
   \begin{table}[t]
      \caption{CER comparison with the representative ASR systems. (with LM)}
      \vspace{6pt}
      \label{table:compare_o}
      \centering
      \scalebox{0.88}{
       \begin{tabular}{l c c c}
        \toprule
        \textbf{Model} &&
        \textbf{Dev}&
        \textbf{Test} \\
        \midrule
        TDNN-Chain (Kaldi) \cite{povey2016purely} && - & 7.45 \\
        SA-T (Transducer) \cite{tian2019self} && 8.30 & 9.30 \\
        LAS \cite{shan2019component} && - & 10.56 \\
        Speech-Transformer \cite{tian2020spike} && 6.57 & 7.37 \\
        SA-Transformer (our implement) && 5.83 & 6.63 \\
        LDSA-Transformer && 5.79 & 6.49 \\
        HA-Transformer && 5.66 & 6.18 \\
        \bottomrule
       \end{tabular}
       }
   \end{table}

\section{Conclusions}\label{sec:4}
  In this paper, we first replaced the common SA in speech recognition by DSA.
  Then, we proposed LDSA to restrict the attention scope of DSA to a local range around the current central frame.
  Finally, we combined LDSA with SA to extract the local and global information simultaneously. Experimental results on Aishell-1 demonstrate that the LDSA-Transformer achieves slightly better performance with lower computational complexity than the SA-Transformer; the HA-Transformer further improves the performance of the LDSA-Transformer; and all proposed methods are significantly better than the three representative ASR systems.

% To start a new column (but not a new page) and help balance the last-page
% column length use \vfill\pagebreak.
% -------------------------------------------------------------------------
\vfill\pagebreak

% References should be produced using the bibtex program from suitable
% BiBTeX files (here: strings, refs, manuals). The IEEEbib.bst bibliography
% style file from IEEE produces unsorted bibliography list.
% -------------------------------------------------------------------------
\footnotesize
\bibliographystyle{IEEEbib}
\bibliography{strings,refs}

\end{document}